\documentclass[sigconf]{acmart}
\usepackage{subfig, caption, enumitem, graphicx}

\AtBeginDocument{%
  \providecommand\BibTeX{{%
    \normalfont B\kern-0.5em{\scshape i\kern-0.25em b}\kern-0.8em\TeX}}}

\copyrightyear{2022}
\acmYear{2022}
\setcopyright{acmcopyright}\acmConference[WWW '22]{Proceedings of the ACM Web Conference 2022}{April 25--29, 2022}{Virtual Event, Lyon, France}
\acmBooktitle{Proceedings of the ACM Web Conference 2022 (WWW '22), April 25--29, 2022, Virtual Event, Lyon, France}
\acmPrice{15.00}
\acmDOI{10.1145/3485447.3512133}
\acmISBN{978-1-4503-9096-5/22/04}

\begin{document}

\title{Characterizing, Detecting, and Predicting Online Ban Evasion}

\author{Manoj Niverthi}
\authornote{Both authors contributed equally to this research.}
\affiliation{%
  \institution{Georgia Institute of Technology}
  \city{Atlanta}
  \state{Georgia}
  \country{USA}
}
\email{manojniverthi@gatech.edu}
\orcid{0000-0003-0147-6759}

\author{Gaurav Verma}
\authornotemark[1]
\affiliation{%
  \institution{Georgia Institute of Technology}
  \city{Atlanta}
  \state{Georgia}
  \country{USA}
}
\email{gverma@gatech.edu}
\orcid{0000-0001-6182-9857}

\author{Srijan Kumar}
\affiliation{%
  \institution{Georgia Institute of Technology}
  \city{Atlanta}
  \state{Georgia}
  \country{USA}
 }
\email{srijan@gatech.edu}
\orcid{0000-0002-5796-3532}

\begin{abstract}

Moderators and automated methods enforce bans on malicious users who engage in disruptive behavior. However, malicious users can easily create a new account to evade such bans. Previous research has focused on other forms of online deception, like the simultaneous operation of multiple accounts by the same entities (sockpuppetry), impersonation of other individuals, and studying the effects of de-platforming individuals and communities. Here we conduct the first data-driven study of ban evasion, i.e., the act of circumventing bans on an online platform, leading to temporally disjoint operation of accounts by the same user. 

We curate a novel dataset of $8,551$ ban evasion pairs (parent, child) identified on Wikipedia and contrast their behavior with benign users and non-evading malicious users. We find that evasion child accounts demonstrate similarities with respect to their banned parent accounts on several behavioral axes --- from similarity in usernames and edited pages to similarity in content added to the platform and its psycholinguistic attributes. We reveal key behavioral attributes of accounts that are likely to evade bans. Based on the insights from the analyses, we train logistic regression classifiers to detect and predict ban evasion at three different points in the ban evasion lifecycle. Results demonstrate the effectiveness of our methods in predicting future evaders ($AUC = 0.78$), early detection of ban evasion ($AUC = 0.85$), and matching child accounts with parent accounts ($MRR = 0.97$). 
Our work can aid moderators by reducing their workload and identifying evasion pairs faster and more efficiently than current manual and heuristic-based approaches. 
\end{abstract}

%
%


%

\maketitle

\vspace{-2mm}
\section{Introduction} \label{sec:1}

As online platforms take a central role in facilitating information sharing and consumption, establishing connections, and enabling discussions~\cite{hampton2011social}, they have also made it easier for malevolent individuals to engage in online abuse~\cite{vidgen2019much, ratkiewicz2011detecting}. 
Existing research has studied the online behavior of malicious users~\cite{kumar2017army, kumar2017antisocial} and several machine learning-based automation tools have been developed to help moderation~\cite{alvari2019less, kumar2017army, solorio2013case}. 
However, malicious users frequently develop new strategies to circumvent detection,  adding more complexities to the entire moderation cycle. Ban evasion is a popular circumvention strategy that malicious users frequently adopt. Ban evasion is the act of circumventing suspensions on an online platform~\cite{RedditBE}, wherein banned users create another account to continue their activities on the platform. In recent years, platforms like Twitter, Reddit, Facebook, Wikipedia, Khan Academy, Discord, Twitch, and eBay~\cite{TwitterBE, RedditBE, WikipediaBE}, have noted how ban evaders threaten the ethos of online platforms by continuing to engage in malicious behavior ranging from harassment~\cite{TwitchHarassment} to spreading terrorist propaganda~\cite{bbcTerror}. Ban evasion has also been linked to real-world acts of mass violence~\cite{gothard2021incel}. Even though the state-of-the-art machine learning models can detect instances of hate speech and incorrect misinformation, they are of limited use as malicious actors can evade the ban and continue disruptive activities. Therefore, it is crucial that we understand the behavior of ban evaders and develop reliable technologies to effectively predict and detect them.

While it may be easy to detect ban evasion on platforms that require personal and sensitive information (email, bank account details, IP address, etc.) while creating an account, like eBay, it is immensely difficult to identify ban evasion when this information is not available to the moderators (such as in cases of subreddit moderators) and on platforms that are not centered around ``real life'' identities --- like most social and information sharing platforms.\footnote{Platforms like Wikipedia do not enforce sharing personal details because, for a few contributors, having their "real life" identity discovered can threaten their ``well-being, careers, or even personal safety''~\cite{WikipediaOuted, bruckner2021inferring}.} Moreover, the current process to identify ban evaders is manual~\cite{WikiSock} and heuristic-based (such as using IP addresses)~\cite{kumar2017army}, which is prone to errors (e.g., two people in the same location can have the identical/similar IP address).
To this end, we analyze the behavioral attributes of ban evaders and develop methods to detect and predict ban evasion based on these attributes, while not relying on sensitive information that is only available for specific platforms.

In this work, we curate a novel dataset of ban evasion in Wikipedia\footnote{Dataset available at \url{https://github.com/srijankr/ban_evasion}} --- a community-driven  encyclopedia that exemplifies the ``free and open'' culture of online communities, and yet struggles with the harmful behavior of malicious users that is also pervasive on other platforms like Reddit, Twitter, and Facebook. Our ban evasion dataset is derived from sockpuppets (i.e., multiple undisclosed accounts operated by the same person) identified and verified by Wikipedia moderators. In all, we identify $8,551$ ban evasion pairs (parent, child) on Wikipedia. We also get their associated account-level metadata (such as creation time, block time, and username) and edit-related information (pages edited, added text, deleted text, edit comment, and timestamp).
We describe the ban evasion lifecycle, and discuss the behavioral attributes (based on meta-information and linguistic signals) that \textit{(a)} characterize future ban evaders from non-evading malicious actors, \textit{(b)} distinguish ban evasion pairs from control malicious pairs, and \textit{(c)} associate evasion child account with the corresponding parent account. Our analyses demonstrate that future evaders differ from other malicious actors in terms of their account-level metadata and linguistic attributes like usernames, the number of edits, use of objective language, and use fewer swear, informal, and sexual words. Furthermore, even though some ban evaders adopt strategies to be deceptive, their holistic behavior, in terms of writing style and temporal signatures, still matches with their previously banned counterparts. 
Based on the ban evasion lifecycle, we formulate three classification tasks: predicting whether an account will evade the ban, early detection of evasion, and matching the evasion accounts with their parent accounts. 
The evaluation of these machine learning-based methods suggests that they are effective in predicting and detecting ban evasion.

Our work can directly aid Wikipedia moderators by \textit{(i)} providing them the likelihood of a malicious account evading a ban in the future, \textit{(ii)} detecting newly created accounts that could be evading bans, and \textit{(iii)} collecting evasion-related evidence to investigate reported malicious accounts. We intend to make the ban evasion dataset available to the community to aid future research and develop actionable tools for Wikipedia moderators. {We discuss the broader perspective and related ethics of this work in Section \ref{sec:7}}.

\vspace{-1mm}
\section{Related Work} \label{sec:2}
\subsection{Sockpuppetry} \label{sec:2.1} 
\citeauthor{kumar2017army} (\citeyear{kumar2017army}) define  sockpuppetry as the act of maintaining and controlling more than one accounts \textit{in conjunction} on an online platform. Sockpuppetry, just like ban evasion, falls into the broader research theme of studying deception in online communities~\cite{tsikerdekis2014online, ott2012estimating}. While earlier works have focused on the deceptive nature of accounts that comprise a sockpuppet group~\cite{zheng2011sockpuppet, liu2016sockpuppet}, \citeauthor{kumar2017army} (\citeyear{kumar2017army}) find that in the context of discussion forums, sockpuppet accounts can be both pretenders (with malicious intents) and non-pretenders (that are overtly visible in the community). Several studies have focused on developing automated tools for detecting sockpuppetry~\cite{solorio2014sockpuppet, solorio2013case, tsikerdekis2014multiple}.
Unlike sockpuppetry, ban evasion involves creating a new account strictly \textit{after} the previous account has been banned for malicious behavior. The difference in the temporal structure – simultaneous handling of accounts for sockpuppetry and sequential handling of accounts for ban evasion, affords new considerations as the evasion account's behavior is now informed by the previous ban and its consequences. Our study focuses explicitly on ban evasion and not sockpuppetry.
 
\subsection{Banning and deplatforming} \label{sec:2.2} Ban evasion has been a problem since the inception of the first online social networks almost 30 years ago~\cite{hackett2019you, dibbell1994rape}. In recent years, however, the problem has become much more pronounced because of the massive scaling of social platforms and the interactions they facilitate. As per Grimmelmann's taxonomy of online moderation, banning is one of the four key techniques that moderators can adopt to avoid disruptions caused by malicious actors ~\cite{grimmelmann2015virtues, jiang2019moderation}. 
Several studies have focused on understanding the impact of bans and deplatforming – be it at an individual level~\cite{cheng2015antisocial, jhaver2021evaluating}, community-level (mass removal of xenophobic communities)~\cite{kor2021alt, van2021deplatformization,  saleem2018aftermath, chandrasekharan2017you, rudas2017understanding}, or platform level (banning of right-wing platforms like Parler from Google's app store)~\cite{aliapoulios2021large, aliapoulios2021early}. These studies find that while bans and deplatforming force a large fraction of users to  abandon the platform, a few users and communities evade the ban and continue malicious activities that cause sustained issues relating to abuse, propaganda, and sometimes even leading to real-life acts of violence. Specifically, \citeauthor{ali2021understanding} (\citeyear{ali2021understanding}) report that users who were banned from Twitter for toxic behavior created accounts on Gab, where they became more active but lost their audience.  
While these studies focus on enforcing bans as a moderation technique and studying its consequences, our study focuses on understanding the behavioral attributes associated with the act of ban evasion and its detection.

Besides academic studies, a few key social media platforms like Reddit and Twitch have recently discussed the development of proprietary automated tools to detect ban evasion~\cite{TwitchonT, RedditPost}. These discussions further reinforce the importance of addressing this problem from a broader perspective.  

\begin{figure}[!t]
    \centering
    \includegraphics[width=0.9\columnwidth]{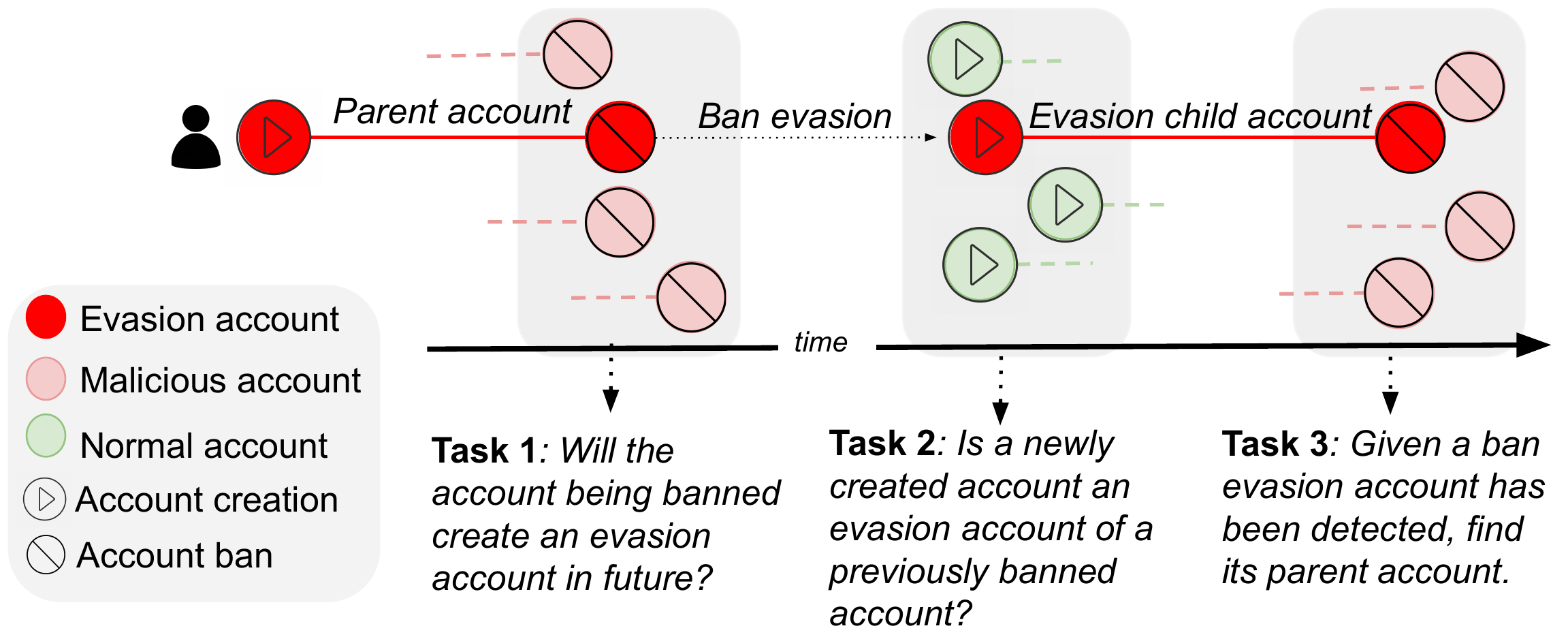}
    \vspace{-1mm}\caption{\textbf{\textit{The ban evasion lifecycle}}. Based on the key stages of ban evasion, we formulate three goals: (i) predict future ban evasion, (ii) detect ban evasion soon after creation of new accounts, and (iii)  detection and matching at the time of banning of evasion child account.}\vspace{-6mm}
    \label{fig:lifecycle}
\end{figure}

\vspace{-2mm}
\section{Ban Evasion Lifecycle and Our Tasks}
\label{sec:lifecycle}

The ban evasion lifecycle (shown in Figure~\ref{fig:lifecycle}) starts with the banning of a malicious account (the ``parent'' account). The ban can be enforced by moderator(s) or an automated algorithm to limit harm to the online community. During evasion, this is followed by the account owner creating a new account (the ``child'' account) to bypass the ban, often to continue abusing community member(s), vandalizing the platform, or engaging in other malicious activities. We refer to this act of creating the child account as \textit{ban evasion}. The child account also engages in malicious behavior and is eventually detected and banned from the platform or community. The three key stages of the ban evasion lifecycle are: banning of parent account, creation of child account, and banning of the child account.

Moderators want to catch evaders early in the evasion lifecycle. 
However, the current process followed by moderators has several shortcomings. First, moderators often fail to report ban evasion activities due to the limited availability of tools that can help them determine whether an account is conducting malicious activities in isolation or is linked to a previously banned account (i.e., whether it is an evasion account). In fact, moderators only identify 10\% of all evaders as per official Reddit statistics~\cite{RedditPost}. Second, the current lifecycle only brings the ban evaders under investigation \textit{after} they have engaged in malicious behavior (which may as well be after multiple instances of abuse and vandalizing). 

We formulate the following tasks, at the three key stages of the ban evasion lifecycle, that can greatly reduce the cognitive load of moderators and will limit the harm done to community members on such platforms:

\noindent $\bullet$ \textbf{Task 1. Evasion Prediction}: Given an account that is banned for malicious behavior, predict if it is likely to create an evasion child account in the future. 

\noindent $\bullet$ \textbf{Task 2. Early Evasion Detection}: Given an account that has been just created, detect if it is an evasion child account of any of the previously banned accounts.

\noindent $\bullet$ \textbf{Task 3. Ban-time Evasion Detection and Attribution}: 

Given an account that has been reported to engage in malicious activity, identify whether it is an evasion child account or an isolated account. If it is an evasion child account, identify its parent account.

\vspace{-2mm}
\section{Wikipedia Ban Evasion Dataset}\label{sec:4}
We first discuss the main dataset that we curated for our study of ban evasion and then describe the construction of negative samples to facilitate comparative analyses.
\vspace{-2mm}
\subsection{Ban Evasion Dataset}
Since there are no existing ban evasion datasets, we curated a new dataset based on Wikipedia's sockpuppet group data.\footnote{\url{https://en.wikipedia.org/wiki/Wikipedia:Sockpuppet_investigations}} A \textit{sockpuppet} is defined as an account that is controlled by a user that also controls at least one other account.
Wikipedia records data regarding instances of sockpuppeteering that have been identified (using a combination of manual and automated techniques) and rigorously verified by moderators. 
As per data collected on March 10, 2021, Wikipedia has identified and banned 19,395 ``sockpuppet groups". Each group contains information about a set of accounts (``sockpuppets") that a single user-controlled. 
We noticed that sockpuppet groups were not all mutually disjoint, so we preprocessed the data by adopting a graph-based approach (please refer to Appendix \ref{app: dataset_details} for details) to create disjoint larger sockpuppet groups that truly comprise all the accounts controlled by one entity (person or organization).
This gave us 18,707 sockpuppet groups. 

There are no explicit labels of ban evasion account pairs. So, we create a novel technique to identify ban evasion accounts from the sockpuppet group data. 
In each sockpuppet group, we first ordered all the accounts temporally by their creation time. 
Then, for each account within a particular group, we identified their temporal predecessor and successor.
An account $u$'s temporal predecessor refers to the account whose ban most closely precedes $u$'s creation. An account $v$'s temporal successor refers to the account created most recently after $v$'s ban.
Then, $(u,v)$ is a ban evasion pair if: (a) both accounts are part of the same sockpuppet group, (b) $u$ is $v$'s temporal predecessor, and (c) $v$ is $u$'s temporal successor. 
This procedure ensures one-to-one mapping between parent and child accounts in a ban evasion pair since multiple accounts could share the same temporal successor (e.g., when one parent creates several evasion accounts) or the same temporal successor (e.g., when multiple accounts were banned and then one evasion account is created).
In the pair, $u$ is the parent, and $v$ is the child account. 
The strict condition and bidirectional criteria of pairing lead to 32,661 ban evasion pairs. 

However, some of the above pairs can be part of the same group. 
In order to avoid instances of repeated ban evasions by a single user, we focus on the \textit{first} ban evasion pair within each group, where this order is defined by the creation time of the parent account within a ban evasion pair. This resulted in the final set of \textit{8,551 ban evasion pairs analyzed in our study}. 

For each of the accounts within these pairs, we collect their contribution history on Wikipedia (meta-information about all revisions, the content of the actual contributions on Wikipedia pages, and comments left on revisions -- see Appendix \ref{app:wiki_details} for details of all the data collected). 
{Note that while it is possible for banned users to create new accounts and not engage in malicious activities thereafter, we specifically focus on instances where the evasion account is also identified and banned because of malicious behavior. The latter is a more acute setting that requires robust and early detection. Thus, our work focuses on the case where both parent and child evasion accounts are malicious. Accordingly, the curated dataset also covers only the appropriate instances. } 

\vspace{-4mm}
\subsection{Task-specific Account Matching}
To characterize the behavior of ban evaders and understand how their behavior differs from benign or other non-evading malicious entities, we need to create task-specific matched samples so that we can contrast their behavior. 

\vspace{1mm}\noindent\textbf{Matched users for Evasion Prediction.}\label{sec:4.2.1} The goal of the Evasion Prediction task is to predict whether a banned parent account will create an evasion account in the future. This task involves comparing two types of accounts known to engage in malicious behavior: (a) banned accounts that will create an evasion account in the future (parent accounts), and (b) malicious users that will \textit{not} create an evasion account in the future. To collect a set of non-evading malicious users accounts, we retrieved a set of $1,354,956$ banned account records from Wikipedia. We then applied a set of filters to this original set, namely removing all accounts that were classified as sockpuppets that violated Wikipedia's proxy policy or were autoblocked, leaving us with $55,027$ non-evading malicious users.

To account for temporal confounds that may influence evasion behavior, we match as per the ban time of a parent account: 
malicious accounts that were banned within a 7-day window of a banned parent account's ban time are matched together.
This way, each parent account is matched to 123.4 non-evading malicious accounts, on average. This process resulted in 1,055,257 total negative samples across all parents.

\vspace{1mm}\noindent\textbf{Matched users for Early Evasion Detection.}\label{sec:4.2.2} The Early Evasion Detection task involves comparing two classes of account \textit{pairs}: (a) true ban evasion pairs consisting of parent accounts and evasion child accounts and (b) matched pairs composed of ban evasion parent accounts and benign Wikipedia accounts who do not engage in malicious activity. We retrieved $10,867,460$ accounts from the set of \textit{all} normal accounts on Wikipedia, removing accounts that make no edits. 
To control temporal confounds, we match users according to their account creation time. 
Specifically, a benign account $u$ is matched to an evasion child account if $u$'s account creation time is within a 1-day window of the child's account creation time and $u$'s creation time was strictly after the parent's ban time. 
Since there is an over-abundance of benign accounts, 
we limited the number of matched benign accounts per child account to 100.
On average, each child account is matched with 41.85 benign accounts, leading to 357,848 negative samples overall.

\vspace{1mm}\noindent\textbf{Matching users for Ban-time Evasion Detection.}\label{sec:4.2.3}
This task involves comparing two classes of account pairs: (a) ban evasion pairs, each consisting of a parent account and a child account, and (b) matched pairs, composed of the same parent account and a matched account that engaged in  malicious behavior without engaging in evasion.
Similar to the previous task, constructing negative samples involved identifying ``candidate" banned accounts for a particular evasion child account. Specifically, non-evading malicious accounts that were created after the parent account's ban time and within a 7-day window of the creation time of the evasion child account were denoted as candidates for false children for a particular false parent. This resulted in 1,003,698 total negative samples across all parents, with an average of 117.4 negative samples per parent. 

\vspace{-2mm}
\section{Ban Evasion Analysis}\label{sec:5}
\vspace{-1mm}
\subsection{How are future ban evaders different from other malicious users?}\label{sec:5.1}
We contrast the behavior of accounts that evade bans in the future (parent accounts) against non-evading malicious accounts, i.e., accounts that do engage in malicious activities and are consequently also banned but do not evade the bans (described in Section \ref{sec:4.2.1}). 

\begin{figure*}[!h]
   \subfloat[]{%
      \includegraphics[width=0.193\textwidth]{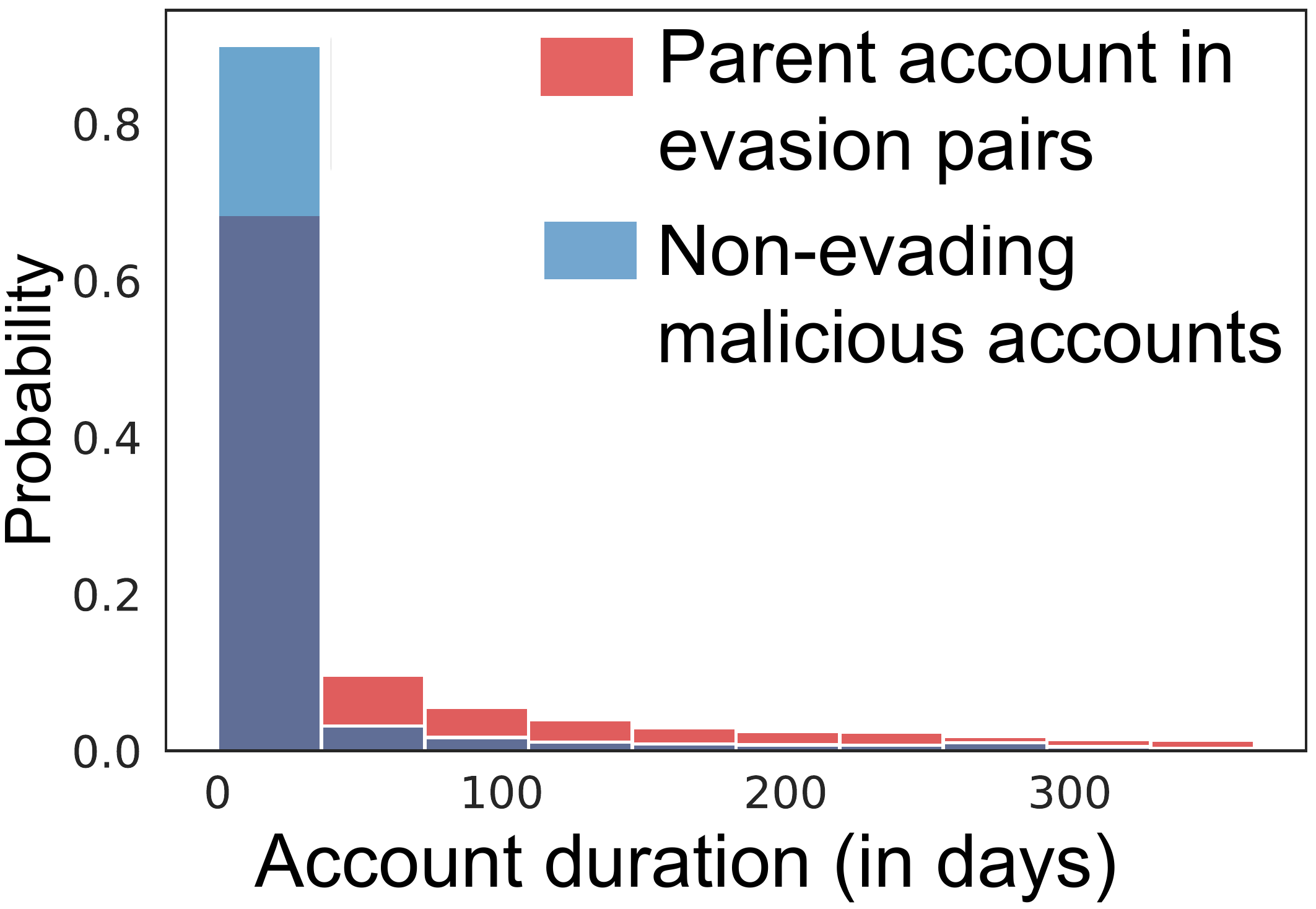}}
\hspace{\fill}
   \subfloat[]{%
      \includegraphics[ width=0.193\textwidth]{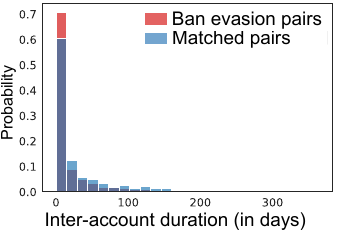}}
\hspace{\fill}
\subfloat[]{%
      \includegraphics[ width=0.193\textwidth]{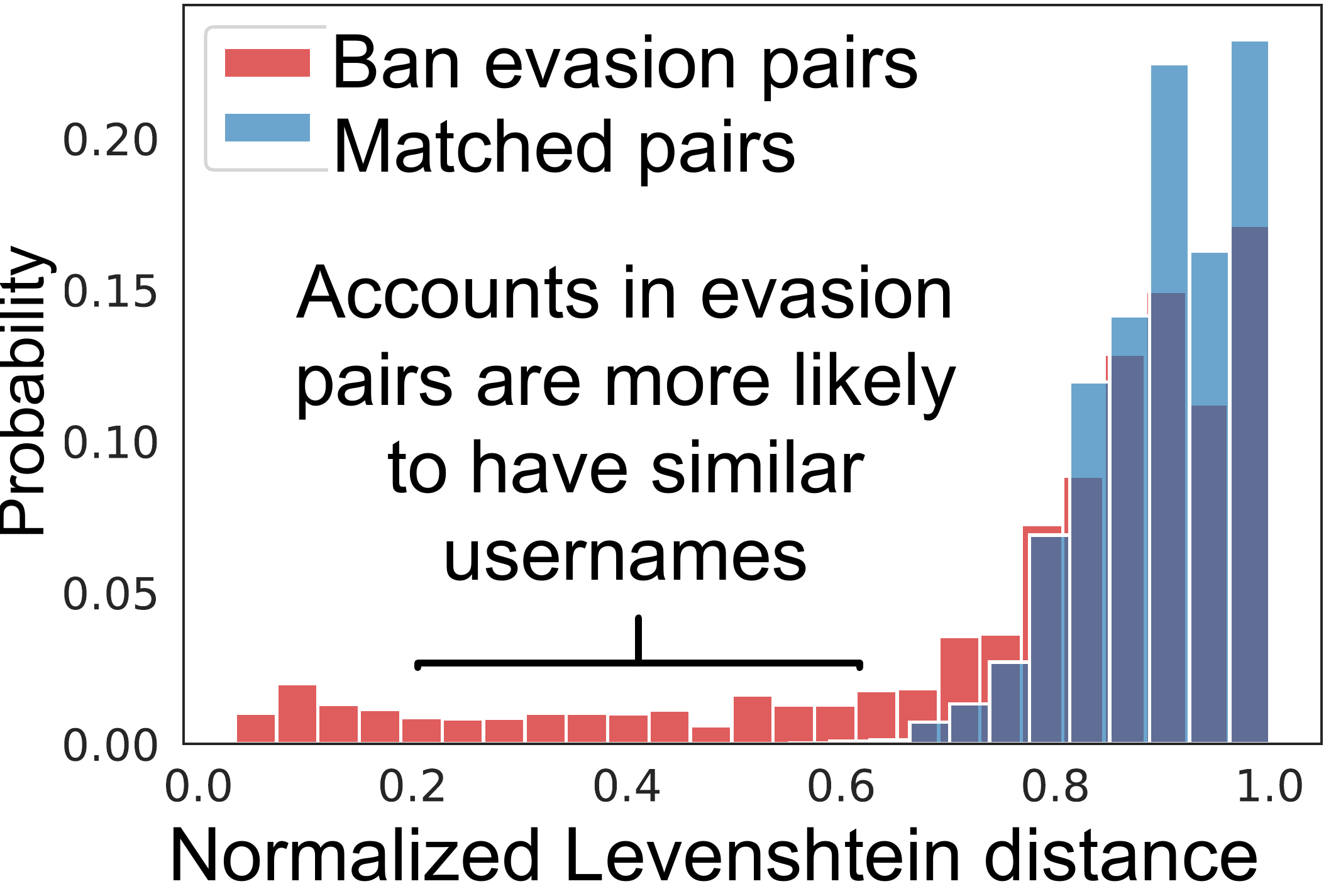}}
\hspace{\fill}
\subfloat[]{%
      \includegraphics[ width=0.20\textwidth]{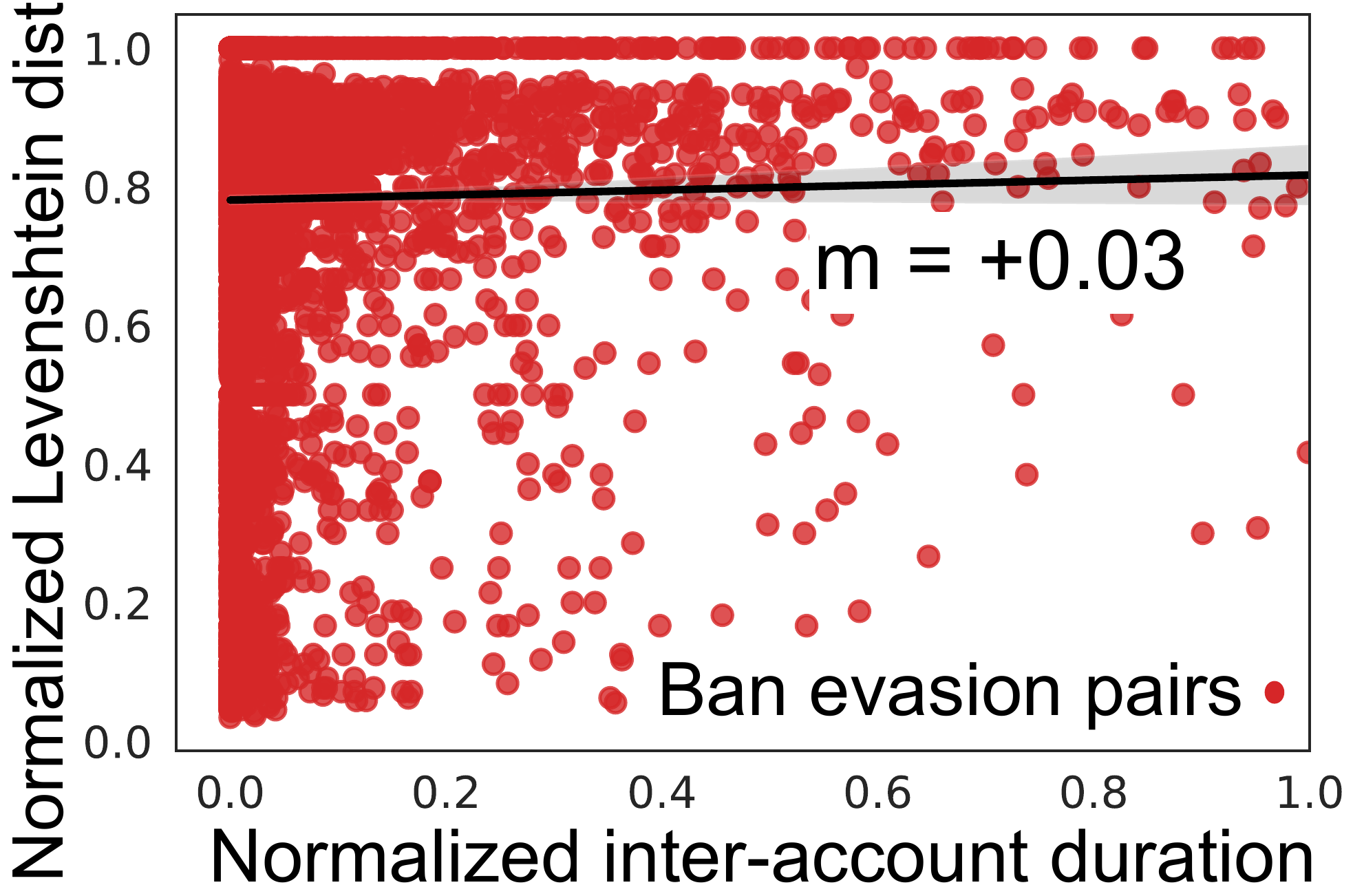}
      }
\hspace{\fill}
   \subfloat[]{%
      \includegraphics[ width=0.193\textwidth]{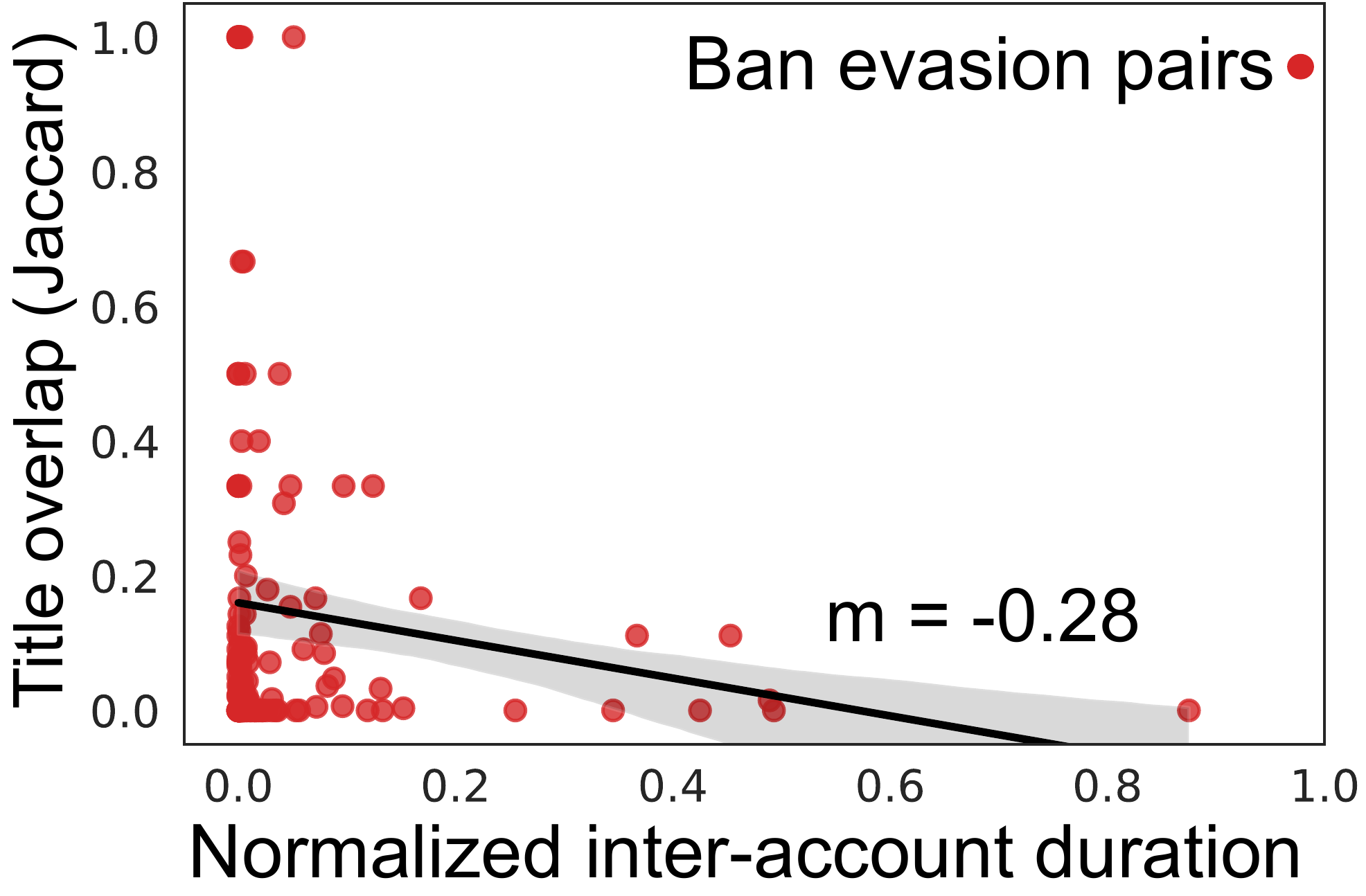}}\vspace{-4mm}
\captionsetup{justification=centering}
\caption{\textit{Ban Evasion Analysis}. \textit{(a)}: Distribution of account duration of ban evasion parents versus non-evading malicious accounts. \textit{(b)}: Distribution of inter-account duration for ban evasion pairs versus matched pairs. \textit{(c)}: Correlation between inter-account duration and normalized Levenshtein distances between usernames. \textit{(d)}: Distribution of inter-account duration for ban evasion pairs. \textit{(e)}: Correlation between inter-account duration and title overlap.}

\label{fig:analysis_plots}\vspace{-3mm}
\end{figure*}

\vspace{1mm}\noindent \textbf{Account duration and activity.} We find that the median duration for which parent accounts are active is about 18 days, during which the median number of revisions that they make is 15 (see Figure \ref{fig:analysis_plots}a). This is considerably higher than the behavior of non-evading malicious accounts -- the median duration for which they are active on the platform is 4 hours, and the median number of revisions is 3. Furthermore, the median number of unique Wikipedia pages that parent accounts edit is higher than non-evading malicious accounts (7 vs. 1). Parent accounts make edits after considerably higher time gaps than their counterparts (median gap of 7 hours vs. 3 minutes). 
Overall, we find that future evaders are active for a longer duration before getting banned. In that duration, they make more revisions, edit more unique pages, and keep a higher separation between their edits than their non-evading counterparts.

\vspace{1mm}\noindent \textbf{Username comparison.}
Account usernames have previously been shown to indicate malicious behavior~\cite{kumar2017army, wang2018online}.
Here, on analyzing the usernames of the parent accounts, we find that, on average, they use shorter usernames in comparison to other malicious accounts
($11.89$ vs. $13.30$ characters; $p < 0.001$)\footnote{all p-values are calculated using Welch's unequal variances $t$-tests.}. A closer view indicates that, on average, parent accounts use fewer non-alphanumeric characters (excluding spaces) ($0.41$ vs. $0.49$ characters; $p < 0.001$). 
Manual inspection (by one author of the paper) of 100 randomly sampled usernames from both sets suggests that non-evading malicious accounts often use non-alphanumeric characters to ``hide'' explicit and inappropriate names (e.g., `A\$\$ KIKR' and `\$hit \$on' (these are real usernames from the platform)). 
Three independent annotators found 1, 2, and 1 instance(s) of inappropriate usernames\footnote{The annotators were instructed to identify whether a username violates Wikipedia's appropriate username policy.} (1 common instance identified by the $3$ annotators) in the set for parent accounts, and  7, 9, and 8 instances of inappropriate accounts (7 common instances across the $3$ annotators) in the set of non-evading malicious accounts, respectively. 
This observation hints at the subdued behavior of future ban evaders --- while some malicious users are offensive and disruptive from the time of account creation, future ban evaders are relatively less obvious.

\vspace{1mm}\noindent \textbf{Text analysis via LIWC.} To contrast the linguistic behavior of these accounts, we leverage the LIWC lexicon to obtain valuable information about emotionality, thinking styles, and attentional focus of the users~\cite{pennebaker2001linguistic, tausczik2010psychological}. Using LIWC, we categorize words into meaningful psycholinguistic categories and represent the text added by each account as the count of words that fall into each of the LIWC categories. We find that parent accounts tend to focus more on the past (frequent use of words like \textit{ago}, \textit{did}, \textit{talked}) than non-evader malicious accounts (0.049 vs. 0.037; $p < 0.05$; Cohen's $d =$ 0.63). Parent accounts also tend to use more function words (0.434 vs. 0.420; $p < 0.05$; Cohen's $d = $ 0.46) and, in particular, more prepositions (0.15 vs. 0.14; $p < 0.01$; Cohen's $d =$  0.38). Additionally, a higher usage of impersonal pronouns (0.022 vs. 0.19; $p < 0.01$; Cohen's $d = $ 0.44) and lesser usage of personal pronouns (0.094 vs. 0.095; $p < 0.05$; Cohen's $d =$ 0.26) could indicate an attempt at demonstrating objectivity in their language. Aligned with the previous finding about subdued malicious behavior, as reflected by fewer instances of inappropriate usernames, we find that parent accounts use fewer swear words (0.0018 vs. 0.0026; $p < 0.05$; Cohen's $d =$ 0.41), fewer informal words (0.0636 vs. 0.0693; $p < 0.05$; Cohen's $d = $ 0.29), fewer affective words (0.0345 vs. 0.0394; $p < 0.01$; Cohen's $d = $ 0.39), and fewer sexual words (0.0012 vs. 0.0024; $p < 0.01$; Cohen's $d = $ 0.54) than non-evading malicious accounts.

Overall, our analyses demonstrate that future ban evaders demonstrate subdued malicious behavior that is less explicit and more camouflaged than non-evading malicious users.

\vspace{-2mm}
\subsection{How are ban evasion pairs different from other malicious pairs?}\label{sec:5.2}

In this section, we consider the pairs of evasion accounts, i.e.,  (parent account, evasion child account), and contrast their traits against pairs that comprise children that engage in malicious behavior but are not evasion accounts, i.e., (parent account, malicious account). These pairs were created as described in Section \ref{sec:4.2.3} earlier, ensuring that the temporal confounds while comparing `evasion child account' against `malicious account' are accounted for.

\vspace{1mm}\noindent\textbf{Account duration and activity}. First, we study the temporal dynamics of ban evasion. Based on the $8,551$ pairs of evasion accounts, we note that the evasion child account is created after a median duration of approximately 4 days and 20 hours, although we observe a high variance in this time gap with a standard deviation of 255 days. The large variation in the duration (see Figure \ref{fig:analysis_plots}b) between the ban of the parent account and the creation of the evasion child account could possibly hint at the varying intentions of the creators~\cite{kumar2017army} – while some may want to join the platform right after their ban to resume their previous activities, others may want to rejoin the platform after a relatively long idle period with different intentions. Once created, the evasion child accounts are active on the platform for a median duration of 3 days and 12 hours, which is significantly shorter than the preceding parent accounts (median $=$ 18 days). During this period, the median of the number of revisions that evasion child accounts make is 6. Again, although the raw number of edits made by the evasion child account is outmatched by those made by the parent account (median $=$ 15), it is important to note that on a per-day basis, child accounts make more revisions. The increased rate of edits made by child accounts may signify additional motivation of a user to proliferate a certain message and reinforce why it is important to detect ban evasion accounts early. 

\vspace{1mm}\noindent \textbf{Username comparison.} Next, we analyze the similarity between the usernames of the parent and child accounts. For this, we compute the normalized Levenshtein distance between the two usernames\footnote{We normalize the Levenshtein distance measure to be $\in [0, 1]$ by dividing the edit distance with the maximum length among the two strings under consideration.}. We find that, on average, the usernames of accounts within an evasion pair are \textit{more} similar when compared the matched pairs $(0.86 \text{ vs. } 0.91, p<0.001)$; see Figure \ref{fig:analysis_plots}c. To further investigate this, we analyze the distribution of normalized Levenshtein distance with respect to the inter-account duration. The inter-account duration is defined as the length of time for which the perpetrator was inactive after the ban of the parent account and was normalized to be $\in [0, 1]$ after removing accounts that had greater than $1,000$ days ($> \mu \pm 2\sigma$: a typical threshold for outlier elimination) of inter-account duration. Figure \ref{fig:analysis_plots}d shows that a considerable number of users form evasion accounts soon after the initial ban with a similar username. 
For some users, even with higher idleness duration, their new usernames are notably similar to the parent accounts. 

\vspace{0.01in}\noindent\textbf{Edit overlap.}
We now focus on the pages that evasion pairs make edits to. For each evasion pair, we consider the set of pages that the parent account edited and the set of pages that the evasion child account edited. We quantify the overlap between these two sets by computing the intersection-over-union (Jaccard index). We find that, compared to the matched pairs, accounts within an evasion pair have a higher overlap in the set of pages that they edit $(0.12 \text{ vs. } 0.0001, p < 0.001)$. However, an interesting aspect to consider is the correlation between the idleness duration before creating the evasion account and the overlap between the edited pages. We find the correlation to be -0.08 for the evasion pairs (see Figure \ref{fig:analysis_plots}e). This observation hints at a relatively higher presence of users who edit the same pages after making their evasion child accounts soon after the parent account is banned, indicating an intent to return to the platform with similar motives. As the inter-account duration increases, the evasion child accounts tend to edit different pages than the parent accounts. 

We also analyze the commonality between the meta-comments that accounts leave as a part of their edits, which are supposed to describe the edits to the moderators and other contributors. We compute the Jaccard index between the unigrams of comments left by the evasion child account and the parent account. We find that in comparison to the matched pairs, accounts within a true pair have greater overlap of unigrams $(0.08 \text { vs. } 0.01; p < 0.001)$. 

\vspace{1mm}\noindent\textbf{Linguistic traits of evasion pairs.}
To capture the linguistic traits of these pairs of accounts, we focus on three aspects of content added on Wikipedia by the accounts within a pair: \textit{(a)} overlap of unigrams, \textit{(b)} semantic similarity, and \textit{(c)} variation in psycholinguistic attributes. As prior studies have demonstrated that sockpuppet accounts demonstrate similarities along many of these behavioral axes~\cite{kumar2017army, solorio2013case}, we assess whether such similarities also exist between accounts within a ban evasion pair, where the goal of evasion child account is precisely not to be identified and matched to the banned parent account.

\vspace{1mm}\noindent\textbf{Unigram overlap.}
Since parent and child evasion accounts edit similar pages, they will likely have a similar vocabulary. To check if this is the case, we quantify their unigram overlap. 
For any given pair of accounts, we compute the Jaccard index (intersection over union) between the set of unique unigrams used by the parent account and the set of unique unigrams used by the evasion child account. We find that evasion pairs have a Jaccard index of $0.118$ (95\% CI: $0.114$, $0.122$) whereas matched pairs have a Jaccard index of $0.046$ (95\% CI: $0.043$, $0.051$). The difference between the Jaccard indices is statistically significant ($p < 0.001$). A larger value of the Jaccard index for the true evasion pair indicates that evasion pairs use similar words when contrasted against matched pairs.

\vspace{1mm}\noindent\textbf{Embedding-based similarity.} While unigram overlap is a trivial measure to compute linguistic similarity, it does not take semantic similarity of language into account. To do so, we compute the embedding-based similarity of the sentences written by parent and child accounts. For each account in a given pair, we first take the average of all BERT~\cite{devlin2018bert} sentence embeddings (by taking the mean of all token embeddings within a sentence) for a given account. Then, we compute the cosine similarity between the averaged sentence embeddings of the parent and child accounts. We find that evasion pairs have an embedding-based similarity of $0.911$ (95\% CI: $0.909$, $0.913$), while matched pairs have an embedding similarity of $0.832$ (95\% CI: $0.830$, $0.833$). 
Adding to the relatively higher value of unigram overlap, a higher semantic similarity for true pairs indicates a substantial similarity between the content that parent and child accounts add. The difference between evasion and matched pairs' embedding-based similarities is statistically significant ($p < 0.001$).

\vspace{1mm}\noindent\textbf{Psycholinguistic analysis.}
The content added to Wikipedia pages itself can map to several psycholinguistic concepts. Using the LIWC lexicon~\cite{pennebaker2001linguistic}, we obtain a psycholinguistic representation of all the text added to the platform for each account. Then, we compare the representation of accounts within a pair by computing the category-wise absolute difference, and then taking an average across all categories. We find that the overall absolute difference for evasion pairs is $0.035$ (95\% CI: $0.034$, $0.036$), while that for matched pairs is $0.039$ (95\% CI: $0.038$, $0.040$). The difference is statistically significant ($p<0.001$). The lower value of average absolute difference for evasion pairs demonstrates that new evasion accounts display similar psycholinguistic attributes while adding content to online platforms as the banned accounts.

\vspace{-3mm}
\subsection{How do evaders change their behavior using the child account?}\label{sec:5.3}

Since ban evasion violates platform policies, evaders are incentivized to adapt their behavior to avoid detection. However, it is not clear how the behavior of evasion child account changes with respect to that of the parent account. To understand these behavioral changes in a psycholinguistic sense, we analyze the category-wise change in LIWC scores between child and parent accounts within an evasion pair. We find that child accounts demonstrate an increased usage of words that indicate cognitive processes (like \textit{cause}, \textit{know}, \textit{ought}) when compared to the previously banned accounts ($0.066$ vs. $0.049$; $p < 0.001$). This could possibly hint at an increased attempt to justify their claims on Wikipedia. Similarly, child accounts increase the usage of words that indicate social processes (like \textit{brother}, \textit{talk}, \textit{friend}) after getting banned ($0.114$ vs. $0.092$; $p < 0.01$). Notably, on average, evasion accounts use a similar number of swear words but show a slight decrease in the use of sexual terms ($p < 0.05$), which could be due to lesser explicitness in their malicious activities. 

\vspace{1mm}\noindent \textbf{What makes a ban evasion attempt successful?}
It is essential to understand why certain evasion attempts are successful while others are not. Thus, we categorize ban evasion pairs into two groups. 
\textit{(i) Successful evasion pairs} comprise evasion pairs in which the child account was active for a longer duration than the parent account. Conversely, \textit{(ii) unsuccessful evasion pairs} comprise evasion pairs in which the child account was active for a lesser duration than the parent account. We note that out of the 8,551 evasion pairs, 2,718 (31.8\%) are considered successful evasion pairs, and 5,833 are considered unsuccessful (68.2\%) evasion pairs. 

One of the most distinguishing psycholinguistic attributes of successful pairs is in their usage of swear words--- 
evasion child accounts in successful pairs slightly decreased the usage of swear words as compared to their parent accounts (change from parent to child use of swear words is -0.007); on the other hand, the evasion child accounts in unsuccessful pairs used swear words slightly more than the parent accounts (difference between parent and child is 0.003). This difference is statistically significant ($p< 0.01$). Additionally, child evasion accounts within successful pairs increase the use of impersonal pronouns (0.021 vs. 0.016; $p < 0.05$) and words that relate to cognitive processes (sharing insights, implying causation) (0.096 vs. 0.053; $p< 0.05$) when compared against their unsuccessful counterparts. This could hint at efforts by successful evaders to appear more objective and logical in their contributions. 
Successful evasion child accounts also have usernames that are more distinct from their parents than unsuccessful evasion child accounts (normalized Levenshtein distance between child's and parent's usernames: 0.91 vs. 0.84; $p < 0.01$). Interestingly, the extent to which successful child accounts edit the same pages edited by the parent accounts is greater than that of unsuccessful child accounts (Jaccard's index: 0.136 vs. 0.115; $p < 0.01$). 

Thus, our analysis indicates that despite continuing to make edits on the same pages, the successful child accounts can evade moderators for a longer duration due to their restrained linguistic attributes and different usernames. 

\vspace{-4mm}
\section{Predicting \& Detecting Ban Evasion}\label{sec:6}
The above analyses provide key insights into the behavior of ban evaders. In this section, we use these insights to develop machine learning-based methods for prediction and detection of ban evasion that can \textit{(i)} inform the moderators with the likelihood of future evasion while banning malicious accounts, \textit{(ii)} help them by flagging suspicious evasion child accounts, and \textit{(iii)} identify pairs of parent and child accounts.

\vspace{1mm}\noindent\textbf{Features.} We use various attributes considered in the analyses above as features for building our classifiers. We divide features into several subcategories: (a) temporal features, (b) edit history-based features, and (c) linguistic features. Tables \ref{tab:feature_table1} and \ref{tab:feature_table2} give an overview of these feature sets. {It is worth noting that while some features like username similarity are good for explanatory analyses~\cite{hofman2021integrating}, they can be manipulated easily, and hence we do not use them in prediction and detection settings. Manipulation of linguistic attributes incurs a relatively higher cognitive cost on ban evaders.}

In each task (described below), we combine features across the three feature sets by adopting recursive feature elimination~\cite{guyon2002gene} and report final results. We also perform ablation studies by training the model with individual subsets of features. To control for randomness, we assess each model's performance 5 times over the same data by setting different random seeds, and report the mean values. {We observe narrow confidence intervals in all the cases.}

\vspace{-2mm}
\subsection{Evasion Prediction (Task 1)} \label{sec:6.1}Our first goal is to predict whether an account that is banned for engaging in malicious activities will create an evasion account in the future. Providing online moderators with the ability to determine which accounts are likely to attempt ban evasion prior to the manifestation of the attempt will allow them to proactively combat and prevent such infractions from occurring, limiting the scale of the abuse across the platform. 
We use the classification features described in Table \ref{tab:feature_table1} and train a logistic regression classifier to perform the classification task.

\vspace{1mm}\noindent\textbf{Training and testing datasets.} 
For this task, we create positive samples as the set of ban evasion parent accounts and negative samples as the set of non-evading malicious accounts, as we described in Section \ref{sec:4.2.1}. We split the set of parent accounts temporally into train and test sets.  
We assigned the first 80\% of the parents 
(ordered by their creation time) to the train set and the remaining parents to the test set. We assigned each parent's negative samples to the same set as the parent. We then removed all overlaps between negative samples across both the train and test sets to avoid information leakage. In all, this gave 15,695 samples in train (1:120 ban evasion parent to non-evading malicious user ratio) and 2,555 samples in test (1:120 ban evasion parent to non-evading malicious user ratio).

\vspace{1mm}\noindent\textbf{Results.} As shown in Figure \ref{fig:rmsd_plots}(A), the classifier achieves an AUC of 0.778.  
This is a high value considering the difficulty in making this future prediction, given only the information about the current malicious account.
Notably, feature ablation studies show that the edit history features provide an AUC of 0.724, providing the strongest signal. Other feature sets alone resulted in near-random performances, with temporal and linguistic features both reporting AUC values of 0.487 and 0.555, respectively. 

\vspace{-1mm}
\subsection{Early Evasion Detection (Task 2)} \label{sec:6.2}
This task aims to deduce, after an account has been created and has made $k$ edits, whether or not the account is an evader account? For experimentation, we chose the value of \textit{k} to be 3, which is about half of the median number of edits that a child evasion account makes before it is identified. Based on the feature set discussed in Table \ref{tab:feature_table2}, we train a logistic regression classifier to distinguish between the evasion children and accounts that are not malicious. Providing online moderators the capability to determine whether newly created accounts are evaders based on their relation to past accounts can aid in fighting instances of ban evasion as early in the ban evasion lifecycle as possible.

\begin{table}[!t]
\small
\resizebox{8cm}{!}{
\begin{tabular}{p{0.18\linewidth} | p{0.75\linewidth}}
\textbf{Feature set}      & \textbf{Features}\\ 
\hline
Temporal & account creation time (day of week, month, and day), account ban time (day of week, month, and day), account duration \\
\hline
Edit history     & number of unique pages account has edited, total number of contributions made, average duration between contributions, average difference per contribution\\
\hline
Linguistic       & average of psycholinguistic scores (LIWC), average in sentiment scores (VADER)
\end{tabular}
}
\caption{\small{Account-level features for prediction.}}\vspace{-9mm}
\label{tab:feature_table1}
\end{table}

\begin{table}[!t]
\small
\resizebox{8cm}{!}{
\begin{tabular}{p{0.18\linewidth} | p{0.75\linewidth}}
\textbf{Feature set }     &  \textbf{Features}\\ 
\hline
Temporal & parent creation time (day of week, month, and day), parent ban time (day of week, month, and day), child creation time (day of week, month, and day), child ban time (day of week, month, and day), duration of parent, duration of child, inter-account duration \\
\hline
Edit history     & overlap in edited pages (Jaccard), overlap in unigrams left in comments (Jaccard)\\
\hline
Linguistic       & unigram overlap (Jaccard), embedding similarity (BERT), absolute difference of psycholinguistic scores (LIWC), absolute difference in sentiment scores (VADER)
\end{tabular}
}
\caption{\small{Pairwise features for detection and ranking.}}\vspace{-9mm}
\label{tab:feature_table2}
\end{table}

\begin{figure*}[!t]\vspace{-1mm}
   \subfloat{%
      \includegraphics[ width=0.17\textwidth]{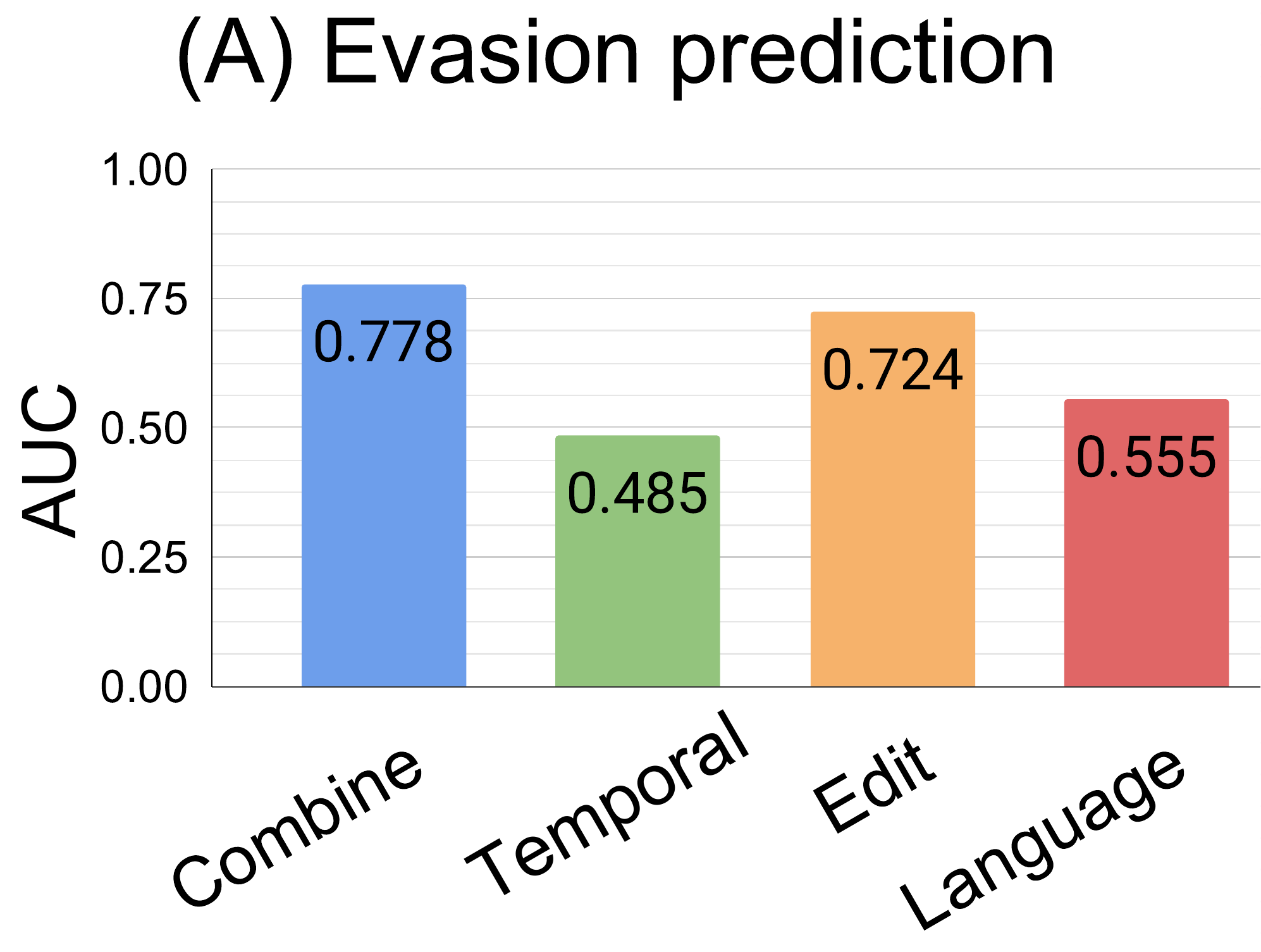}}
\hspace{3mm}
   \subfloat{%
      \includegraphics[ width=0.19\textwidth]{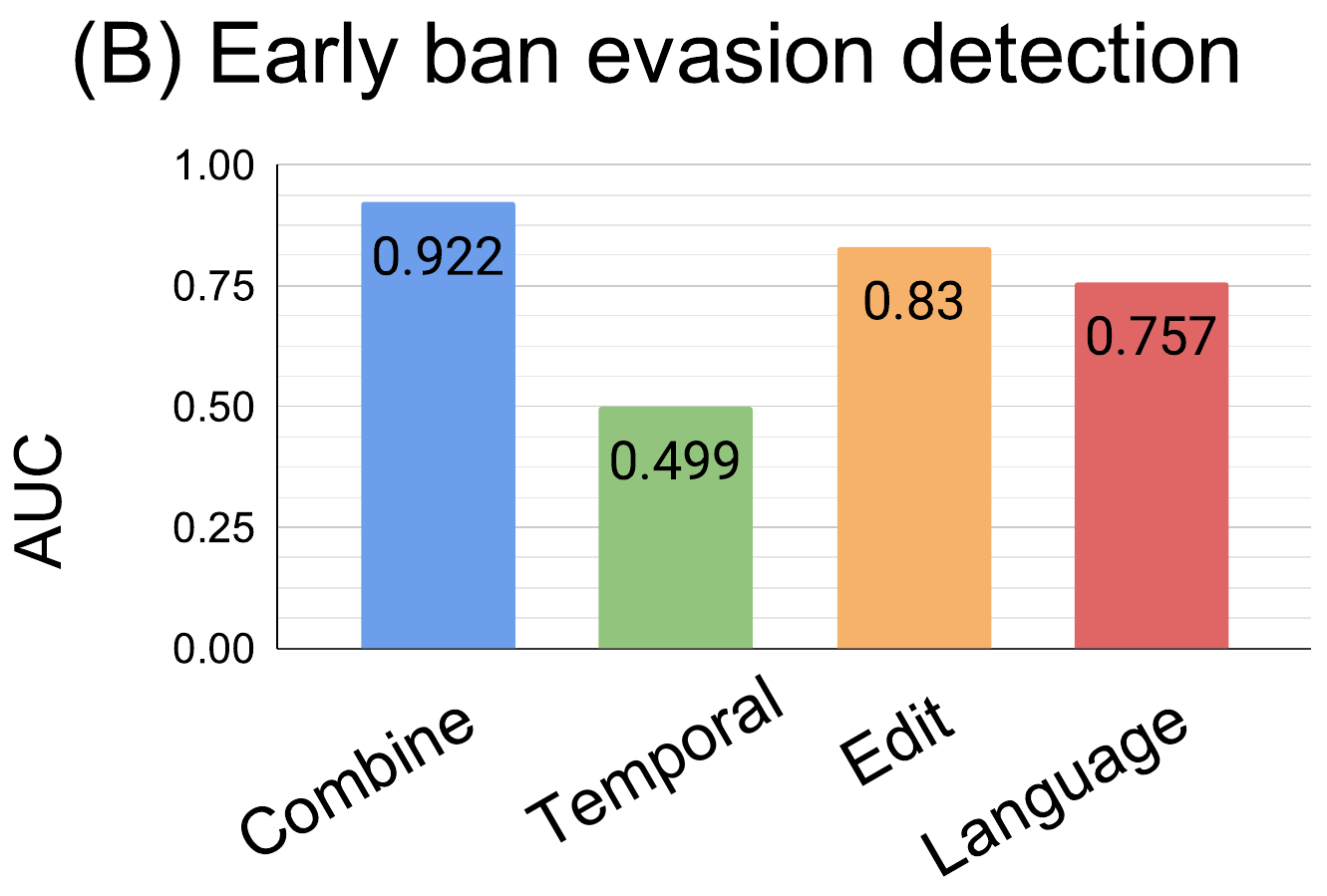}}
\hspace{3mm}
   \subfloat{%
      \includegraphics[ width=0.173\textwidth]{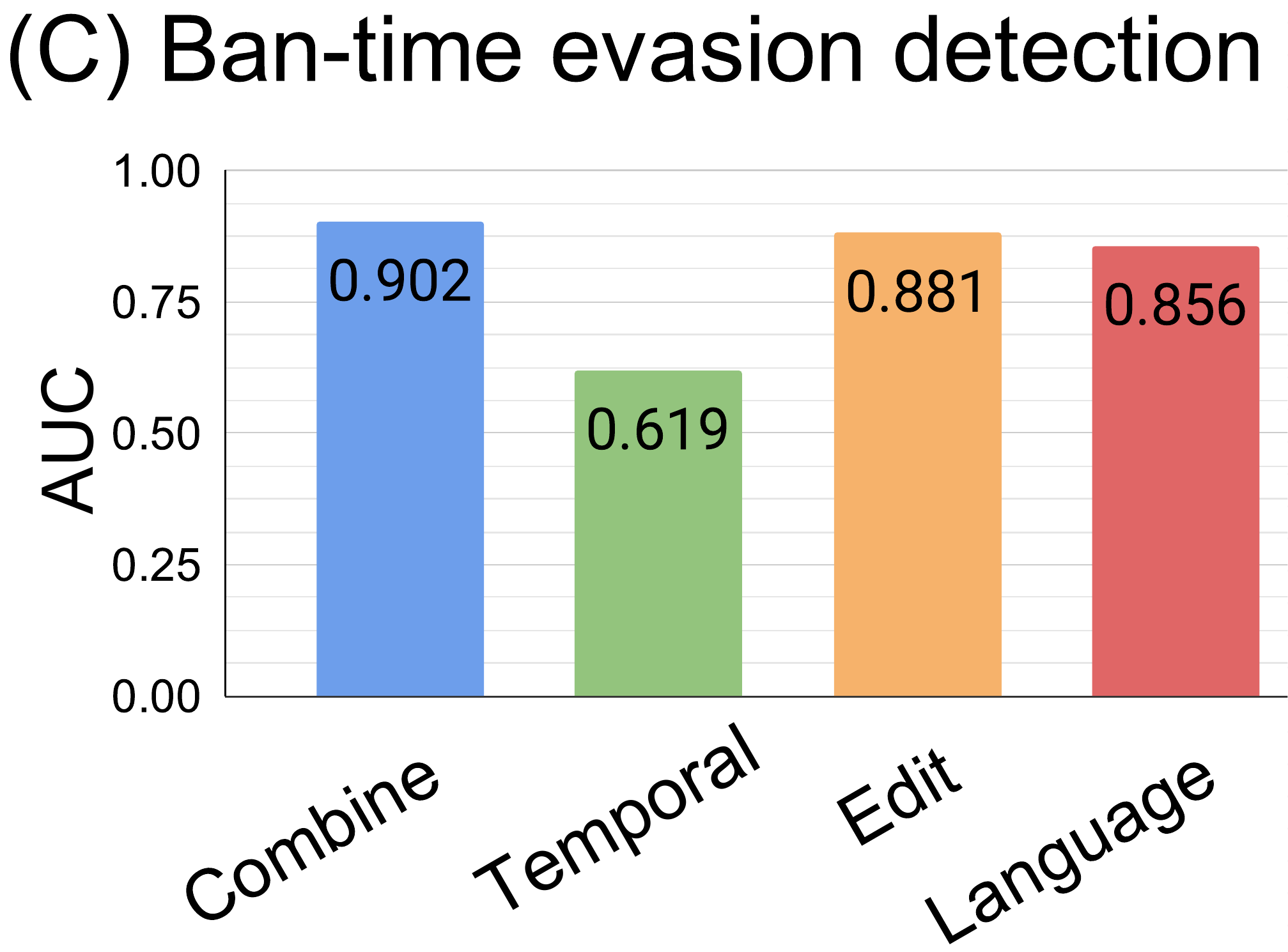}}
      \hspace{3mm}
    \subfloat{%
      \includegraphics[ width=0.17\textwidth]{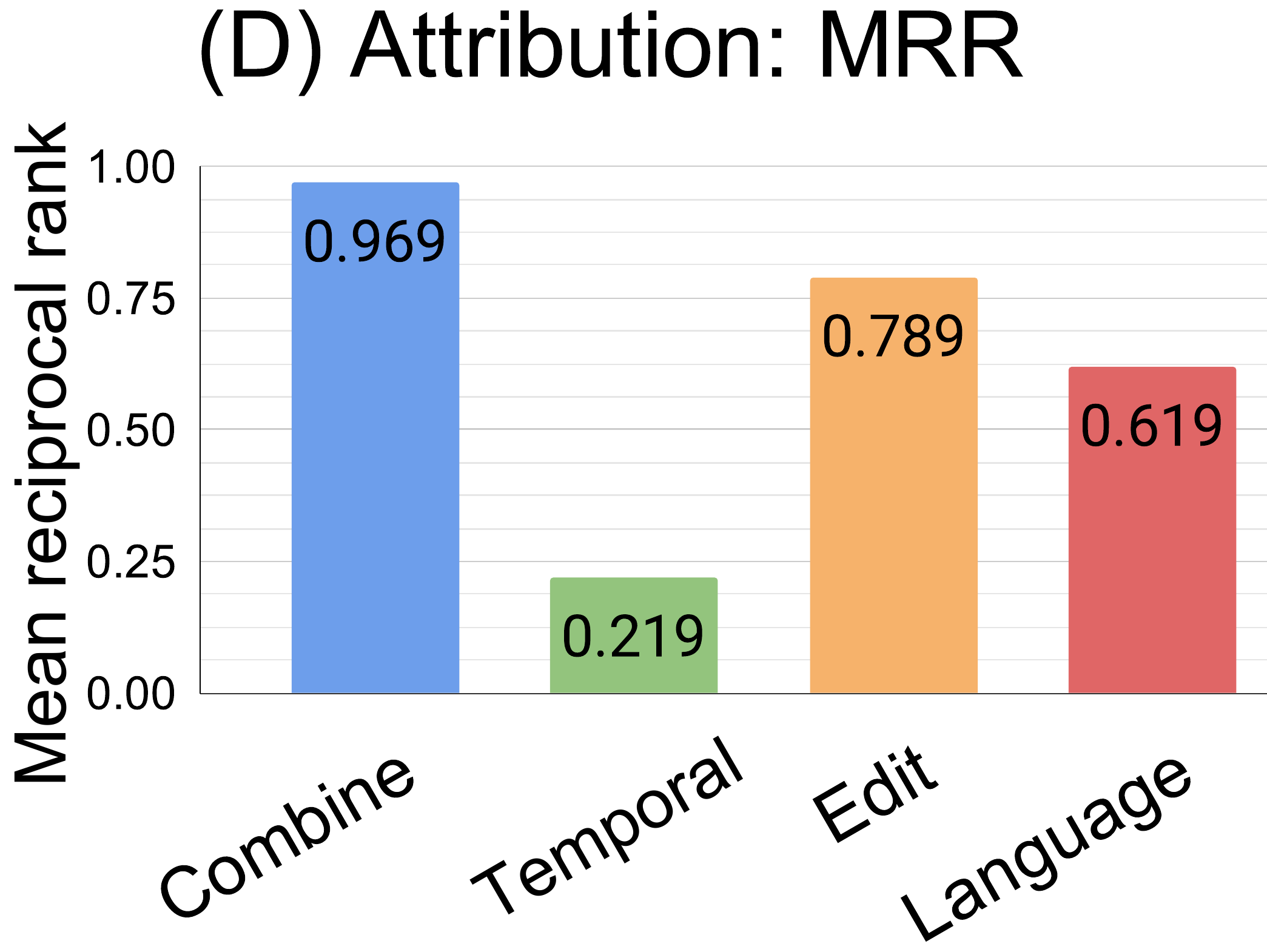}}
    \hspace{3mm}
    \subfloat{%
      \includegraphics[ width=0.173\textwidth]{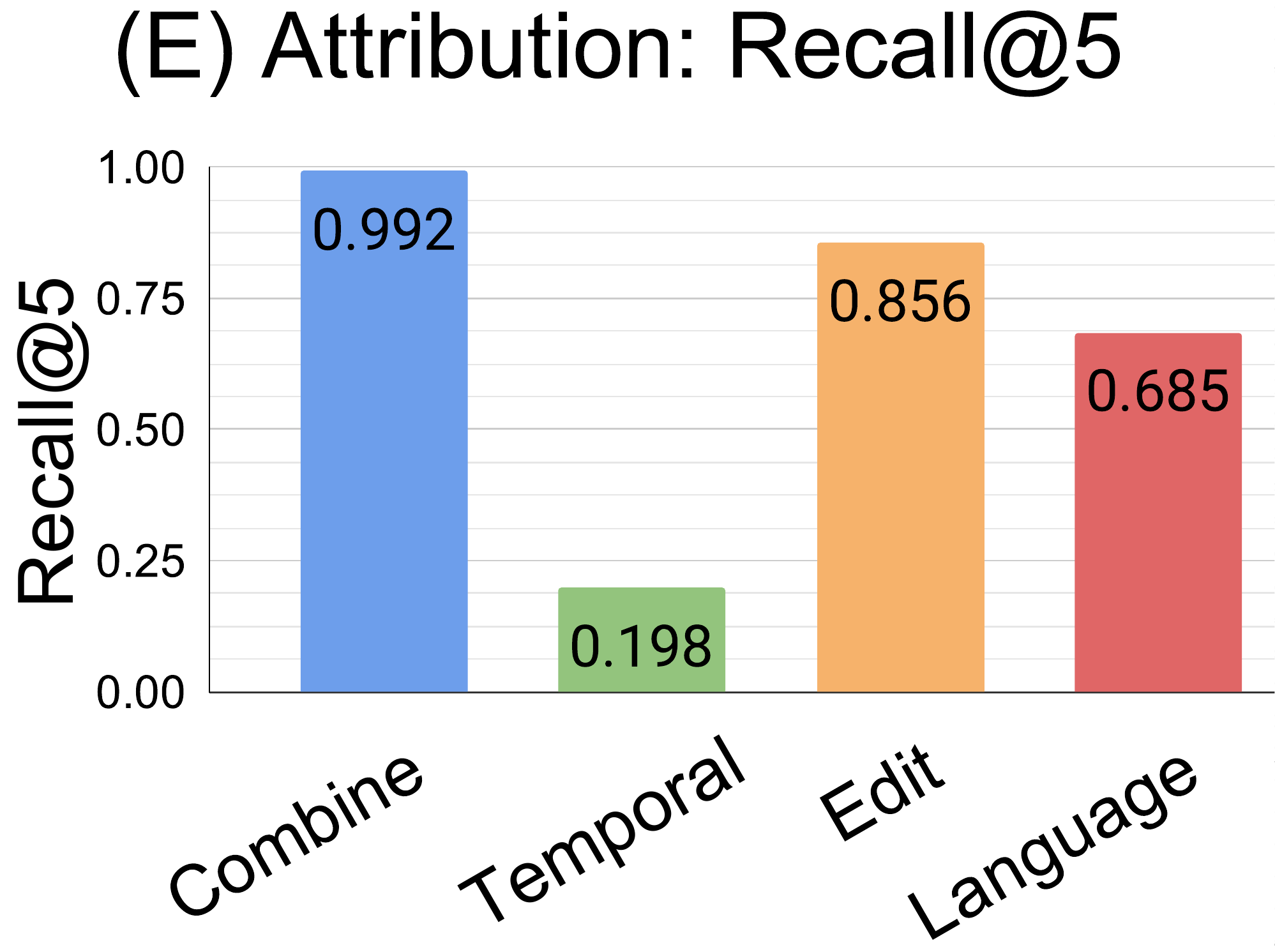}}
      \vspace{-2mm}
      \captionsetup{justification=centering}
\caption{\textit{Performance on the detection, prediction, and attribution tasks}. \textit{(A, B, C)}: The results demonstrate overall AUC after combining all the possible features, as well as using only specific type of features (temporal, edit, and language). \textit{(D, E)}: Attribution of detection evasion child account to the correct parent (MRR  and Recall@5).
}
\label{fig:rmsd_plots}\vspace{-3mm}
\end{figure*}

\vspace{1mm}\noindent\textbf{Training and testing datasets.} 
We use the positive sample set of true ban evasion pairs and the negative sample set of ban evasion parents paired with matched benign users, as described in Section \ref{sec:4.2.2}.
We temporally split the entire set of ban evasion pairs into train and test sets using a chronological ordering by parent account creation time, assigning the first 90\% of ban evasion pairs to the train set and remaining pairs to the test set. We assigned a negative sample to the same train or test set as its corresponding positive sample. 
We removed overlaps between negative children assigned to train and test to prevent leakage. This resulted in 216,516 samples in train ($1 : 38.08$ ban evasion pair to matched pair ratio) and 34,706 samples in test ($1 : 56.84$ ban evasion pair to matched pair ratio). 

\vspace{1mm}\noindent\textbf{Detection results.} The model provided high separability between the two classes, resulting in an AUC of 0.853 when all features were included. Similar to the previous task, the temporal feature set gave a relatively poor performance, with an AUC of 0.499. The edit and linguistic feature sets also performed similarly, with their AUC scores presented in Figure \ref{fig:rmsd_plots}(B). This shows that the model can correctly identify ban evasion child accounts soon after creation, thereby being useful to  moderators to prevent damage. 

\vspace{-2mm}\subsection{Ban-time Evasion Detection and Attribution (Task 3)}\label{sec:6.3}
This task has both a detection and attribution component. In the detection task, we want to detect given an account that has been identified as malicious, whether it is an evasion account or an independent malicious account. In the attribution task, given a known evasion child account, we want to match it with its parent account. This task is crucial for its potential to aid human moderators in collecting the evidence to support and inform their resultant penalty -- more penalty for evading account. We train and test a logistic regression classifier on the feature set detailed in Table \ref{tab:feature_table2} using the train and test split described below.

\vspace{1mm}\noindent\textbf{Training and testing datasets.} \label{sec:task3datasplit}
For the task, we utilized ban evasion pairs as our positive samples and matched pairs as our negative samples (ban evasion parents matched with malicious accounts that did not attempt evasion), as described in Section \ref{sec:4.2.3}. Subsequently, we temporally split the set of ban evasion pairs temporally, assigning the first 90\% of pairs to the train set, as ordered by the creation time of parent accounts. We then assigned negative samples to the groups based on the presence of their respective parent account in either of the two sets. All overlaps between negative samples assigned to train and test sets were eliminated. This resulted in 241,858 samples in train ($1:14.7$ ban evasion pair to matched pair ratio) and 90,293 samples in test ($1:15.1$ ban evasion pair to matched pair ratio). 

\vspace{1mm}\noindent\textbf{Detection results.} Overall, the model exhibited high distinguishing performance among evaders versus non-evaders. The model had a high AUC of 0.902 when all features were included (see Figure \ref{fig:rmsd_plots}(C)). {To evaluate the detection capabilities of our model for both successful evaders as well as unsuccessful evaders (see Section \ref{sec:5.3}), we conduct a fragmented evaluation. We observe that the same model identifies successful ban evaders from malicious non-evaders with 0.909 AUC and unsuccessful ban evaders from malicious non-ban evaders with 0.898 AUC.}

\vspace{1mm}\noindent\textbf{Matching candidate parents for ranking task.}\label{sec:matching_data} The above binary classification task aims to distinguish evasion accounts from other isolated malicious accounts. However, once identified, evasion accounts need to be matched with their parents. To this end, we formulate a ranking task: given a child evasion account, rank its true parent account higher than other candidate parents. 

We create the following dataset for the ranking task. For a given child account $c$, candidate parents include its true parent and other recently-banned parents that were (a) not its true parent, and (b) were banned before $c$'s creation. 
The temporal constraints to identify candidate parents are added because we have seen that most child evasion accounts are created shortly after the ban of the parent (as shown in Section \ref{sec:5.2}). We add at most 50 recently-banned parents to the candidate set. 
The data is then split into train and test by doing a temporal split --- first 90\% child accounts in the train set and the rest in the test set. Candidate accounts are matched with their respective child accounts. 
This resulted in an  train set with 230,651 accounts and a test set of 19,215 accounts, with an average of 49.9 candidate parents per child.

We train a logistic regression classifier that distinguishes between ban evasion pairs and mismatched pairs. 
Once trained, we rank all the potential parents for a child based on the positive class probability score of the trained classifier. Finally, the rank of the true parent is calculated in the ranked list of all candidate parents. 

\vspace{1mm}\noindent\textbf{Ranking results.} We evaluate the matching performance using Mean Reciprocal Rank (MRR) and Recall@K. MRR computes the reciprocal rank of the true parent account in the ranking, averaged over all test instances. A higher value of MRR indicates better ranking capabilities. Similarly, in our case, Recall@K reports the fraction of test instances in which the  correct parent account was in the top-K ranks. As shown in Figure \ref{fig:rmsd_plots}(D and E), when considering all features, we see an MRR of 0.969 and Recall@5 value of 0.992,  demonstrating that our model can rank the true parent first among the candidates in many cases, and in the top-5 in almost all cases. 

\vspace{-3mm}
\section{Discussion and Conclusion}\label{sec:7}

This study provides the first understanding of the behavior of ban evasion account pairs in online communities while also providing a practical methodology for predicting and detecting these pairs as they operate online. As moderators struggle to cope with ban evasion, the need to operationalize detection methods beyond human review is more pertinent than ever. By providing online platforms with a generalized framework and tooling for understanding ban evasion, moderators will have an easier time not only in detecting instances of ban evasion but also providing evidence for their penalties~\cite{matias2019civic}.
Additionally, since online discourse is leveraged to inform important policies regarding public health and democracy, among others, it is important to weed out bad actors and only focus on the opinions of benign users. 

\vspace{0.03in}\noindent{\textbf{Broader perspective and ethics.} It is important to note that such evasion prediction and detection tools must be used with caution: moderators cannot penalize someone purely based on suspicion of ban evasion but must still remain vigilant to its occurrence on online platforms. We advocate that such tools should only be used in \textit{egregious} instances of ban evasion and \textit{supplement} human moderators. Completely relying on automated and overly stringent approaches can lead to raised barriers to entry for newcomers~\cite{halfaker2013rise} and systematic propagation of inequities on online platforms~\cite{tripodi2021ms}. In an effort to make online communities healthier for all participants, this work studies the behavior of malicious ban evaders specifically and develops approaches to aid human moderators.}

\vspace{0.05in} It should be noted that even though our work is focused on ban evasion on Wikipedia, many of the methods and features discussed are not Wikipedia-specific. For instance, the attributes quantified using LIWC can be used to identify psycholinguistic similarities in posts on other social media platforms, such as Reddit, Twitter, or Facebook.
Similarly, content-based similarity measures, similarity in usernames, and activity-based measures can inform mitigation efforts across platforms. We encourage interested researchers and practitioners to extend our study to other platforms to develop complementary insights about ban evasion.
The current work focuses on ban evasion pairs in which the child account is also malicious, and thus, also banned. It will be important to study how accounts improve after being banned. 
Furthermore, ban evasion can be iterative, where the users who have evaded bans previously make other attempts to evade the ban of their evasion accounts. 
We intend to study the change in behavior of evaders over multiple rounds of evasion as part of future work. It is also possible that some evaders operate more than one evasion accounts simultaneously after getting banned. While the current study focuses on one parent-child pairs, it will also be interesting to explore cases where multiple parents or child accounts are present. 

\vspace{0.05in}\noindent\textbf{Acknowledgements:} {This research is supported in part by NSF IIS-2027689, NSF OIA-2137724, Georgia Institute of Technology, IDEaS, Facebook,  Adobe, and Microsoft Azure. We thank members of Wikimedia Foundation for discussions, Rohit Sridhar and Kartik Sharma for helping with manual inspections of account usernames, and the anonymous reviewers for their constructive comments.}

\bibliographystyle{ACM-Reference-Format}
\bibliography{main}

\appendix

\section{Appendix}
\subsection{Details about dataset curation}
\label{app: dataset_details}
To reconcile the overlaps between different groups, we constructed a graph where the vertices represented users, and an undirected edge existed between two users in the graph if a single user, utilizing one of these accounts as their primary account, was known to have a false persona they were manipulating through the other account. This graph had 157,067 unique vertices and 158,106 edges. Using a connected components algorithm on this graph, we identified and constructed 18,707 unique groups such that there were no account overlaps between pairs of groups, rectifying the overlap issue present in the original dataset.

After constructing these unique groups, we ordered all the accounts in each group temporally by their creation time. The master account of each group was defined as the first account created in each group. This temporal ordering also provided a method of constructing ban evasion pairs. 
For each account within a particular group, we indexed the account's temporal predecessor and temporal successor. An account's temporal predecessor refers to the account whose ban most closely precedes the given account's creation. An account's temporal successor refers to the account that is created most recently after the given account's ban. Using both these mappings simultaneously, we define a ban evasion pair as a pair of accounts such that two criteria are met: \textit{(a)} the pair's parent account must have the child account as its temporal successor and \textit{(b)} the pair's child account must have the parent account as its temporal predecessor. The bidirectional nature of this pairing ensures one-to-one correspondence, providing us with 32,661 ban evasion pairs. 

\subsection{Contributions Allowed on Wikipedia}
\label{app:wiki_details}
Wikipedia provides the following key functionalities to contributors: (i) user talks --- user can leave messages on a specific user's talk page, (ii) contribution to Wikipedia pages --- users make edits on Wikipedia pages, and (iii) contribution comments --- users contextualize or explain their edits by leaving comments for other contributors and/or moderators. For this work,  we primarily focus on the content of the actual contributions that users make on Wikipedia pages and the comments they leave ((ii) and (iii) above).

\end{document}